# Developing the Reliable Shallow Supervised Learning for Thermal Comfort using ASHRAE RP-884 and ASHRAE Global Thermal Comfort Database II

Kanisius Karyono, *Student Member, IEEE,* Badr M. Abdullah, Alison J. Cotgrave, Ana Bras, and Jeff Cullen

**Abstract**—The artificial intelligence (AI) system designer for thermal comfort faces insufficient data recorded from the current user or overfitting due to unreliable training data. This work introduces the reliable data set for training the AI subsystem for thermal comfort. This paper presents the control algorithm based on shallow supervised learning, which is simple enough to be implemented in the Internet of Things (IoT) system for residential usage using ASHRAE RP-884 and ASHRAE Global Thermal Comfort Database II. No training data for thermal comfort is available as reliable as this dataset, but the direct use of this data can lead to overfitting. This work offers the algorithm for data filtering and semantic data augmentation for the ASHRAE database for the supervised learning process. Overfitting always becomes a problem due to the psychological aspect involved in the thermal comfort decision. The method to check the AI system based on the psychrometric chart against overfitting is presented. This paper also assesses the most important parameters needed to achieve human thermal comfort. This method can support the development of reinforced learning for thermal comfort.

———————————— ◆ ————————————

## 1 INTRODUCTION

THE decarbonising heat and buildings has become one of the vital commitments of many countries, including the United Kingdom (UK), as the response to the UN COP26 summit in Glasgow, 2021. The UK is committed to reaching net-zero emissions by 2050 [1]. The support includes the incentive to decarbonise homes and funding new ideas and technologies. The government will also increase the effort to reduce dependence on burning natural gas in homes and introduce a Future Homes Standard by 2025 with low carbon heating and world-leading levels of energy efficiency [2]. This effort is significant because 86.9% of our time is spent indoors [3], even before the pandemic.

The heat pump is becoming more common to be installed to lower the carbon footprint for residential heating. The UK government will commit to deploying 600,000 heat pumps every year by 2028 [4]. However, implementing the heat pump is not a drop-in replacement for gas-boilers [5]. Prior to the installations, an assessment should be done to optimise the heat pump usage and ensure efficiency.

If the heat pump is installed in poorly performed or leaky buildings, the efficiency will decrease. Efficiency improvements can be made, for example, by adding internal or external insulation and improvements to fabric airtightness [6]. In the UK, 53.36% of houses were built prior to 1964s' which usually have poor performance in heating compared to modern houses, which are better insulated [7].

Using the other types of electric heating with intelligent control will be crucial to fill in the gap while still maintaining the energy efficiency target. The infrared, radiant heater for a specific heating area, a ceramic heater to accommodate the off-peak loads or other kinds of heaters that give an additional benefit are becoming crucial. Electric heating will be among the technology considered to be carbon-friendly. This trend highlights the importance of the research on thermal comfort even before 1920 [8].

There is two significant research in thermal comfort, the human physiology approach and the human psychology/ adaptive behavioural approach [8]. Each approach has benefits and limitations. There has been much research to develop each approach until now, but combining the two approaches is not an easy job. Due to the advancement in artificial intelligence (AI) technology, combining both approaches is now open. This work aims to elaborate on the human physiology and adaptive approach to harvesting their benefit for better energy conservation whilst maintaining human comfort.

This work also proposes new validation methods to

————————————————

- *Kanisius Karyono is with the Electrical Engineering Department, Universitas Multimedia Nusantara, Scientia Garden Gading Serpong, Indonesia, 15810. E-mail: karyono@umn.ac.id, k.karyono@2019.ljmu.ac.uk*
- *Badr M. Abdullah is with the School of Civil Engineering and Built Environment, Liverpool John Moores University, Byrom Street Campus, Liverpool, United Kingdom, L3 3AF. E-mail: B.M.Abdullah@ljmu.ac.uk*
- *Alison J. Cotgrave is with the School of Civil Engineering and Built Environment, Liverpool John Moores University, Byrom Street Campus, Liverpool, United Kingdom, L3 3AF. E-mail: A.J.Cotgrave@ljmu.ac.uk*
- *Ana Bras is with the School of Civil Engineering and Built Environment, Liverpool John Moores University, Byrom Street Campus, Liverpool, United Kingdom, L3 3AF. E-mail: A.M.ArmadaBras@ljmu.ac.uk*
- *Jeff Cullen is with the School of Civil Engineering and Built Environment, Liverpool John Moores University, Byrom Street Campus, Liverpool, United Kingdom, L3 3AF. E-mail: J.D.Cullen@ljmu.ac.uk*

check the learning process in the AI system for thermal comfort based on the psychrometric chart. This validation is crucial to avoid the overfitting problems and minimise the need to use the explainable AI, which is not simple to be deployed in the Internet of Things (IoT) based system and still requires much research. Auburn's Nguyen even mentions that the field of explainability is getting somewhat stuck [9].

## 2 PRELIMINARY RESEARCH
### 2.1 Thermal Comfort
Thermal comfort is a state of mind expressing satisfactory adaptation with the immediate thermal environment. Fanger has set up the milestone model for thermal comfort in 1970 [10]. This work is based on the measurement of human physiology and uses the climate chamber for the measurement. In this chamber, the indoor condition can be simulated and controlled. This model is formulated with the relations among the affecting parameters (1). The approach then expanded with the Predicted Mean Vote (PMV)/Predicted Percentage of Dissatisfied (PPD). This method becomes a standard reference work on thermal comfort, the basis of the ISO 7730–2005 [11] and acknowledged by the ASHRAE-55 Standard [12].

$$M-W = C + R + E + (C_{res} + E_{res}) + S \quad (1)$$

Where: M: the metabolic rate of the occupant
W: mechanical work done by the occupant
C: convective heat loss from the clothed body
R: radiative heat loss from the clothed body
E: evaporative heat loss from the clothed body
$C_{res}$: convective heat loss from respiration
$E_{res}$: evaporative heat loss from respiration
S: the rate at which heat is stored in the body tissues

Since thermal comfort is about a state of mind and Fanger's trial was done in the chamber with limited subjects, it could not represent the different variations in the subject preferences, especially the person with special needs and the difference in the dwelling types. Nicol and Humphreys propose the adaptive method to overcome the problem [14]. The adaptation can be physiological, which is related to the body adaptation to the temperature change, psychological that is formed by the previous human experiences and behaviour related adaptation [15]. Comfort is achieved if people have sufficient opportunities to adapt [16]. The ASHRAE-55 Standard has also acknowledged this adaptive method [17].

Deploying the adaptive thermal comfort models is often done through the black box approach [18]. This deployment makes the model incompatible and cannot be precisely calculated as in the thermal physiology models. Complete thermal adaptation also requires a more extended period. If the training data is taken directly from the occupants during the system implementation, there will not be enough data to cover the whole extreme condition, putting the user in an uncomfortable situation and triggering the user's scepticism. This user experience can negatively impact the system because comfort is about the state of mind [19].

This paper becomes the answer to bridging the two approaches by using the physiology measurement data to train the adaptive model's AI subsystem to accommodate the human behavioural factor. The physiology calculation and psychrometric mapping of the comfort range are used again to verify the result of the learning.

### 2.2 The Possibility of Widening the Comfort Zone
Thermal comfort plays an essential role in the design process of the indoor artificial climate due to its significant impact on health and safety. Productivity is also affected by the thermal condition [20]. The comfort zone can be predicted using the PMV-PPD method [17].

The PMV-PPD comfort zone is included in the ASHRAE-55 Standard [12]. Besides PMV-PDD, the comfort zone is also defined by Givoni [13]. The PMV-PPD is prescriptive, based on the measurement in the thermal chamber. Due to the nature that the comfort is based on satisfactory human adaptation, there is the potential of widening the comfort zone based on the adaptive approach, which is a goal-based method.

The specific group of people like young, elder, disabled, or temporary ill people can have a different comfort range than defined in the PMV-PPD method. For example, the elder people group tend to have higher comfort temperatures [21], [22]. On the other hand, there can be a request for a lower comfort temperature due to the health recommendation. It has been proven that frequent cold exposures can help people have less body fat [23]. Our previous research also shows that the lower temperature setpoint can lower the carbon footprint while maintaining the indoor humidity within the healthy range [24]. Regular exposure to cold acclimation will improve the subjective responses to cold [25]. In the long periods, this might alter the thermal preferences of the occupants.

The comfort zones are shown in Fig. 1. The boundary presents the prescriptive approach, and the number represents the adaptive potential comfort zone. The comfort zone of PMV-PPD is presented as data1 for activity value of 1.0<met<1.3 and clothing value of 1.0 Clo. Data2 present the PMV-PPD comfort zone for 1.0<met<1.3 and 0.5 Clo. Data3-6 represent Givoni comfort zone. Data3 represents the still air condition for winter, data4 for still air condition for summer, data5 for airspeed about two m/s in winter

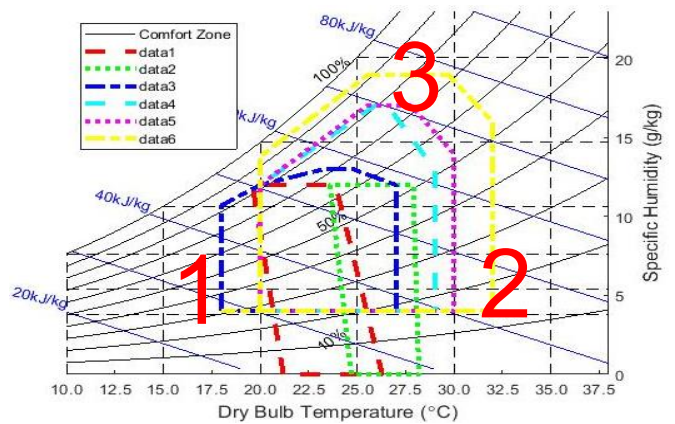

Fig. 1. Psychometrics chart showing the comfort zone [12], [13] and the potential comfort zone [8].

and data6 for summer as seen in Fig.1.

## 2.3 ASHRAE Data

This work uses the most comprehensive thermal comfort dataset, ASHRAE RP-884 and ASHRAE Global Thermal Comfort Database II. These databases are available online as open-source databases. The ASHRAE RP-884 consists of 25,616 entries, and ASHRAE Global Thermal Comfort Database II includes 81,967 entries [26]. This dataset consists of objective indoor climatic observations and subjective questionnaire-based evaluations. The data were taken from the field experiments, with the people doing their activities. Even the data captured the PMV-PPD values; these values differed from the Fanger experiments done in a controlled indoor environment of a climate chamber [27].

This dataset will benefit the AI system developer rather than getting the value based on the own data gathering. Doing the own data collection will require the calibrated sensors equipment, the subject consent and awareness of the thermal sensation and subject questionnaire. The data collection also should be done in the different environmental conditions and building types to get the broad combination of training data. Controlling all of these parameters are challenging in field studies. On the contrary, measuring these parameters in the climate chamber will be easier but not represent the building types and the actual occupants' conditions. There is also the approach to use the simulated or generated data, but the validation of the data can trigger another hurdle.

The ASHRAE database consists of different cities and countries, seasons, climate zone, building type, cooling and heating strategy, and personal information about the subjects. This personal information includes sex, age, height, and weight. Other subjective essential factors are thermal sensation, acceptability, and preference. These subjective factors are taken with specific metabolic rate (met) and clothing insulation level (clo). The comfort indices such as PMV, PPD and Standard Effective Temperature (SET) were calculated uniformly and included in the database. The parameter measurements included in this dataset are various types of temperatures, air velocity, relative humidity and monthly average temperatures. Some indoor environmental controls include blinds, fans, operable windows, doors, and heaters [27].

This dataset has developed many approaches to predict the Thermal Sensation (PTS) [28] regarding the location. Another recent model that is proposed based on this database is done to assess the PMV-PPD accuracy against the database [29]. This work concludes that accuracy varied enormously between ventilation, building types and climate. The authors have seen this gap to propose better models using the power of AI.

## 2.4 Elaborating ASHRAE Data for Supervised Learning

In supervised learning, the role of the training data is crucial. Getting the field data during the system initialisation is not practical because the amount and variety of data will not be adequate. This approach will not give a comfortable experience for the user. In order to accommodate this, the previous study has been done in developing the intelligent system using the previous ASHRAE database.

Previous studies have shown that the AI approach using the ASHRAE RRP-884 has limited diversity and unbalanced distribution. This unbalanced distribution results in a model which is not reliable in extreme conditions [30]. This work uses 20,954 data entries from 25,616 entries after data cleansing. The other research uses ASHRAE Global Thermal Comfort Database II, which consists of a more significant amount of data [31]. The other works use the combination of both ASHRAE RP-884 and ASHRAE Global Thermal Comfort Database II [32], [33]. However, in the previous research, the data items are being selected to represent each class or label. This selection is because the more extensive data does not guarantee higher accuracy and can cause overfitting issues [31].

Not all data from this dataset can be directly elaborated in the training data. This limitation is due to the nature of the human psychological factor that the thermal comfort is personal or individual [32]. The ambiguity of the data inconsistency can be high. The previous work considers this data illogical and an anomaly [33]. From the 107,583 entries, this work only uses 16,795 based on four thermal metrics. The work by Luo also populates only 10,618 entries out of 81,967 entries [31]. This work uses the thermal sensation vote (TSV) for the label of the learning target. Eighty percent of the data are allocated for training and 20% for testing. This work also mentions that allocating more data percentages for testing can decrease accuracy, indicating the lack of data available for training.

The work by Luo discovered that even with 66,3% maximum accuracy using the Radom Forest, the approach already got 10–20% higher prediction accuracy than the PMV model. The model also got 60–66% for 3-point TSV accuracy and 52–57% for 7-point TSV prediction [31]. This work also introduced the six most influencing variables: temperature, relative humidity, clothing insulation, airflow speed, subject age, and activity/metabolism level. The work from Wang also has a similar result. The scalability and sample number are mentioned as limitations, although the accuracy can be increased to 87% [33]. The accuracy in predicting thermal acceptability is also higher than thermal preference.

Learning from the previous work based on the ASHRAE Database, this paper proposes the use of the database which are:
- based on the TSV for the learning target, where three labels for the thermal conditions are used (no change, prefer warmer and prefer cooler)
- uses as much entries data as possible and does not pick pointing data to represent as many individual variations or preferences as possible
- compare the use of four and five most significant variables, which are temperature, relative humidity, clothing insulation, activity/metabolism level, and subject age
- uses simple filtering to minimise the ambiguity of data by considering the psychological aspect of human

Although the ASHRAE dataset is the most comprehen-sive, using the whole dataset for training data is not popular due to the psychological factors present in the data, decreasing the accuracy of the result. This work fills the data conditioning gap to prepare the data to become the training data. This work also uses three state TSV, which will simplify the result to control the heating or cooling (no change, warmer and cooler).

## 2.5 Semantic Augmentation

Getting the data for thermal comfort training is not easy. It requires the proper instruments and consent from the occupants. Most of the entries in the ASHRAE database falls under the "no change" label (43,441 entries). Only about 14,966 entries need a warmer temperature, and about 27,093 entries need a cooler indoor temperature. This case is similar to the image processing and classification problem with highly imbalanced data. The training data for supervised methods are usually difficult to collect due to the costly human efforts and particular domain expertise.

Data augmentation strategy is introduced to balance the feature space to enrich the supervision. The augmentation strategy can normalise the supervision process to improve the robustness by embedding such that the features of the same instance under different augmentations should be invariant and forcefully separated from the other instance features [34].

The previous work shows that data augmentation can be more powerful in the image classification problem if the class identity is preserved, for example, with semantic transformations. Each class in the training set can be added with the samples from the generator. The procedure is computationally intensive and lengthens the training procedure. The training set data can be effectively augmented by searching the semantic directions. The random directions that may result in the meaningless transformation can be reduced [35].

This work aims to develop data augmentation using the approach of semantic data augmentation. The class "no change" remains untouched while the "warmer" and "cooler" classes are added with the data in the semantic direction of the value that is not covered by the ASHRAE database. The "warmer" class is augmented with the lower temperature value under the value of mapped ASHRAE data. On the contrary, the "cooler" class is augmented with the data, which is higher than the mapped ASHRAE data. This method helps to balance the feature space to enrich the supervision. The benefit of this method is that the data obtained from the ASHRAE database is unaffected due to the non overlapped semantic augmentation direction. In this case, the data related to the psychological aspects are still maintained, and the essence of using the ASHRAE database is sustained.

## 3 RESEARCH METHODOLOGY

This work focused on the shallow learning AI for controlling, for example, the electric radiant panel to be deployed as part of the Internet of Things (IoT) system for the residential house. This work is focused on the three TSV values or the thermal preference (no change, warmer and cooler). The diagram that shows the methodology of this work is shown in Fig. 2.

The ASHRAE database is used for the training data source, and the filtering process is implied to maintain the data consistency without eliminating the psychological aspect variations to the data. The semantic augmentation process is added to the data to balance the feature and enrich supervision learning. This work will implement four and five parameters from six dominant parameters due to the availability of IoT sensors.

This work is also proposed to check the AI learning result using a psychrometrics chart. Testing the learning process using the testing data will not be sufficient to check the training result. Visual result validation with the psychrometric chart using a predefined input range to map the thermal comfort zone will result in a higher confidence level of the system. Further, with this visual validation, the parameters can be mapped to view the characteristic of each parameter regarding the impact on thermal comfort. The human psychological/behavioural aspects can be shown on this map.

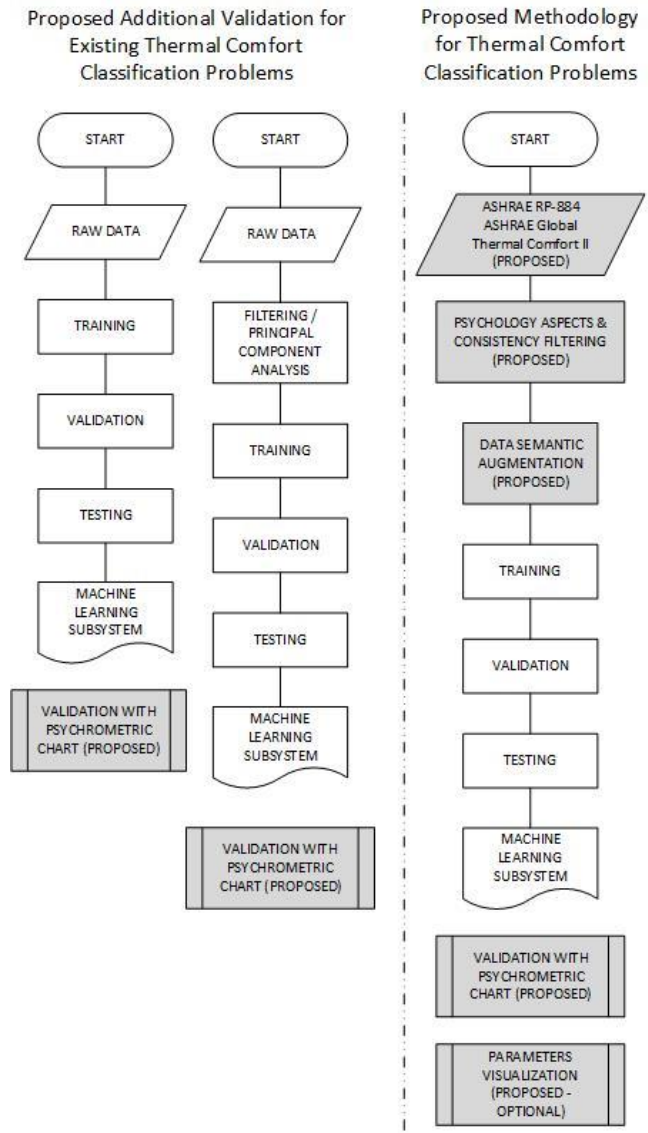

Fig. 2. Proposed validation methods and proposed methodology for thermal comfort AI training

## 3.1 Data Filtering

Previous works that use the ASHRAE database did not use the complete entries but selected entries based on each class to achieve the balanced features. The data used for training was only less than 20% of the whole ASHRAE database. This selection makes the human psychological aspects not easily captured by the supervised training. On the other hand, the risk of overfitting is also implied in using this database. That is why this database is the most reliable for thermal comfort. Since it was published, few AI developers have been willing to use this data to develop their AI systems.

This work proposes simple yet powerful methods to filter the data based on human perception consistency. The need for filtering is because the data was based on precise measurement, but the human perception data was based on the questionnaire that was more prone to error and subjective judgment than the measured data. This filter worked based on the comparison of parameters and omitted the data that is considered to be inconsistent as follows:
1. The filtering is based on the consistency between the thermal acceptability (0 unaccepted 1 accepted) and thermal preference (warmer, no change and cooler). The warmer and cooler should have the thermal acceptability value of 0, no change should have the value of 1.
2. The filtering is based on the consistency between the thermal acceptability (0 unaccepted 1 accepted) and thermal sensation (-3 to +3). The value between -2 to 2 should have the thermal acceptability value of 1, and the other should have 0.
3. The filtering is based on the consistency between the thermal sensation (-3 to +3) and thermal preference (warmer, no change and cooler). The thermal sensation value less than -2 should correspond with warmer, and more than 2 should associate with cooler. The state of no change should have a value between -2 to +2.
4. The filtering is based on the consistency between the thermal preference (warmer, no change and cooler) and thermal comfort (1-very uncomfortable, 6-very comfortable). The value 1-very uncomfortable should not be having the value of no change. The value of 6-very comfortable should not have the value of warmer or cooler.
5. The filtering is based on the consistency between the thermal sensation (-3 to +3) and thermal comfort (1-very uncomfortable, 6-very comfortable). The 1- very uncomfortable value should not have the value of -2 to +2. The thermal comfort value of 6-very comfortable should not have the thermal sensation value below -2 or more than 2.

The target labels for the AI are based on the three states of TSV, -1 means that the occupants need a warmer indoor environment, 1 means that the occupants need a cooler temperature and 0 means that the occupants are satisfied with the indoor temperature. This approach is the most straightforward arrangement for the subject because they are still comfortable in the current temperature, need a lower temperature, or need a warmer temperature.

Not many people can define their thermal preferences using seven scale levels. There is no crisp border between each scale, and even the same temperature can be mapped into the different seven scale TSV. The border between these three scales is not crisp either. However, this map will better understand people's thermal sensations due to its simplicity.

Previous research shows that there are six dominant parameters, and compared to the complete twelve parameters, it only increases 2.6% in the performance compared to elaborating six dominant parameters [31]. The IoT low-cost sensors can detect two of six dominant parameters: temperature and relative humidity. The occupant data entries can introduce the clothing insulation, metabolic rate or activities, and age. Low-cost sensors do not easily detect air velocity. This work also tried to narrow the parameters into five parameters for easier deployment with a residential IoT system. The precise air velocity sensor and the sensor placement will not be feasible for the residential IoT system. This work will give the overview that even without the air velocity sensor, the result of the AI will still be acceptable.

Furthermore, the system which omits the parameter of age is also explored. The purpose of this is due to the high availability percentage for age unavailability in the ASHRAE database. The missing data for the five dominant parameters in the ASHRAE database can be seen in Appendix A.

## 3.2 Data Semantic Augmentation

The base for semantic augmentation is the temperature data. This data is chosen because the class that needs the augmentations are "warmer" and "cooler" classes. Fig. 3 shows the map of both classes, the augmentation and the data map. The class "no change" remains untouched due to the adequate data. This method also retains the psychological aspects and the accurate measurements in the ASHRAE database.

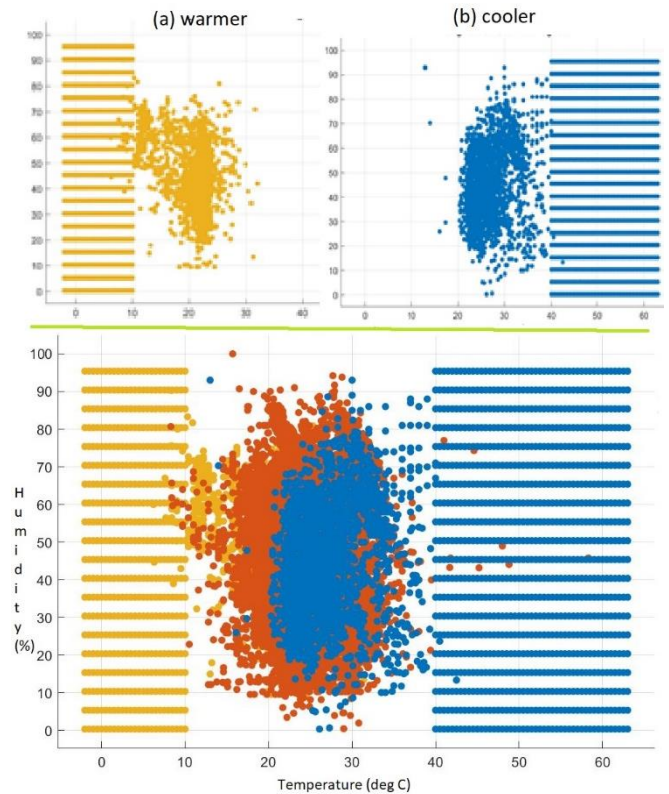

Fig. 3. Database map after the filtering process and semantic augmentation, (a) warmer class (b) cooler class

The augmentation data range was decided based on the notion that the augmentation data will not change the original data obtained from the ASHRAE database. The "warmer" class is augmented with the lower temperature value under the value of mapped ASHRAE data, which is 10 degrees Celcius. The "cooler" class is augmented with the data higher than 40 degrees Celcius. It is shown that the semantic augmentation direction is non overlapped. The essence of the ASHRAE database is sustained. It is shown in Fig. 1. that this range of the augmented data is also outside the comfort zone. The algorithm for generating the semantic augmentation is as follows:

---

**Algorithm 1**: Semantic Augmentation Data

---

```
//%"for colder augmentation class"
Input: data intervals
row=0;
for clo=0 to 2.89 step clo_intervals do
   for met=0.65 to 6.83 step met_intervals do
      for tem=40 to 63.2 step tem_intervals do
         for RH=0.4 to 100 step RH_intervals do
            for age= 6 to 99 step age_intervals do
               row=row+1;
               AugMat(row,1)=clo;
               AugMat(row,2)=met;
               AugMat(row,3)=tem;
               AugMat(row,4)=RH;
               AugMat(row,5)=age;
               AugMat(row,6)="colder";
            end for
         end for
      end for
   end for
end for
Output: Augmentation Matrix: AugMat(row,[1:6])
//%"for warmer augmentation class"
// similar with warmer except this line:
//     for tem=0 to 10 step tem_intervals do
//
//             AugMat(row,6)="warmer";
// temp can be expanded for more extreme temperature
```

---

### 3.3 Psychrometric Based Verification and Parameter Visualization

Testing and validation for supervision learning are usually done using the data fraction randomly populated. The typical value for testing and validation can be 10% to 20%. The more the percentage can lead to the accuracy decrease [31]. However, more extensive data does not guarantee higher accuracy and can cause overfitting issues.

Checking the learning against overfitting issues is not easy. This work proposes using psychrometric chart mapping to validate the supervised learning result. This method is based on the comfort zone map in the psychrometric chart. The overfitting results will lead to the map not showing the correct pattern if the system is fed with the data series. The example of the mapping result can be seen in Fig 4.

The pattern is generated using the validation data generated with the value range of relative humidity from 0% to 100% and temperature from 10 °C to 40°C. Other parameters like age, clothing insulation and activity can be predefined with median values. The result can be mapped with different colours or symbols to represent the output. The blue colour in the sample represents the class that needs warmer temperatures. The red colour represents the class that needs a cooler temperature, while the green represents the comfort zone (no change). In the case of overfitting, the pattern generated will show the pattern very different compared to the comfort zone shown in Fig.1.

This verification process method can also be used to compare the effects of the parameter change. One parameter value can be altered while the other parameters are constant. The impact of each parameter on the comfort zone can be captured and simulated. This method can simplify the representation of the multidimensional parameters that impact thermal comfort.

## 4 RESULT AND DISCUSSION

The aim for data filtering is to use the data entries from the ASHRAE database as much as possible by removing the inconsistent data while still capturing the psychological aspects of human comfort that is registered in the database. This filtered data can be used as a base training data so that the AI developer does not have to capture their data which needs much effort or will decrease the occupants' comfort. The user can then override the setting of the system with their personal preferences. Their personal preferences can be entered in the later stage of the system development.

The filtering consists of five inconsistency checks, and the number of data entries filtered in each filtering item in the ASHRAE database can be seen in Appendix B. The ASHRAE RP-884 database has 25,616 entries, and the ASHRAE thermal comfort database II (1995 – 2015) has 81,967 entries. After the filtering process, the amount of data in ASHRAE RRP-884 is 14,970, with filtered entries of

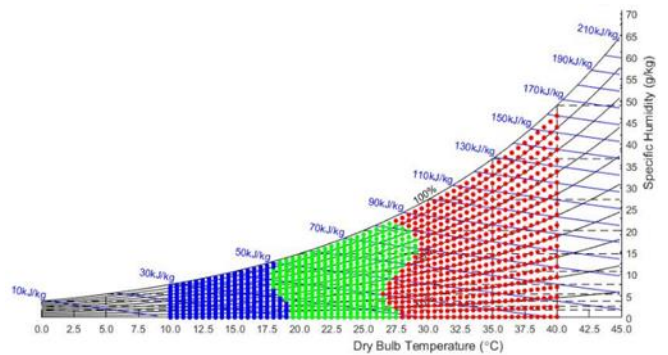

Fig. 4. Mapping the comfort zone generated by the pre-trained system.

10,646 or 41.56%. The ASHRAE Thermal Comfort Database II entries are 50,286 after the filtering process, with the filtered value of 31,681 entries or 38.65%. In total the ASHRAE database after filtering is 65,256 entries or 60.66% (filtered value is 42,327 entries or 39.34%). This entry has at least three times as much data as the previous work. More elaborated data means that the system can better capture the occupants' variations. The risk of overfitting can be eliminated with the further processing the data. The simple algorithm for filtering is as follows:

---

**Algorithm 2**: Simple Data Filtering for the ASHRAE Database

---

//% simple filtering based on five inconsistency check
// entry will be marked as 0 to be filtered and 1 to retain
// the field with "NA" entries will be skipped
**Input:** ASHRAE database
**for** ctr=1 to size (ASHRAE database) **do**

   //%ThermalAcceptability vs ThermalPreference
IF((THERMAL_ACCEPTABILITY=1)*(THERMAL_PREFERENCE="no change"),1,IF((THERMAL_ACCEPTABIITY=0)*(THERMAL_PREFERENCE="cooler"),1,IF((THERMAL_ACCEPTABILITY=0)*(THERMAL_PREFERENCE="warmer"),1,IF(OR((THERMAL_ACCEPTABILITY="NA"),(THERMAL_PREFERENCE="NA")),"NA",0))))

   //%ThermalSensation vs ThermalAcceptability
IF((THERMAL_ACCEPTABILITY=1)*(ABS(THERMAL_SENSATION)>2),0,IF((THERMAL_ACCEPTABILITY=0)*(ABS(THERMAL_SENSATION)<=1),0,IF(OR((THERMAL_ACCEPTABILITY="NA"),(THERMAL_SENSATION="NA")),"NA",1)))

   //%ThermalSensation vs ThermalPreference
IF((THERMAL_PREFERENCE="warmer")*(THERMAL_SENSATION<-2),1,IF((THERMAL_PREFERENCE="cooler")*(THERMAL_SENSATION>2),1,IF((THERMAL_PREFERENCE="no change")*(ABS(THERMAL_SENSATION)<=1),1,IF( (THERMAL_SENSATION>1)*OR((THERMAL_PREFERENCE="cooler"),(THERMAL_PREFERENCE="no change"))*(THERMAL_SENSATION<2),1,IF( (THERMAL_SENSATION<-1)*OR((THERMAL_PREFERENCE="warmer"),(THERMAL_PREFERENCE="no change"))*(THERMAL_SENSATION>-2),1,IF(OR((THERMAL_SENSATION="NA"),(THERMAL_PREFERENCE="NA")),"NA",0))))))

   //%ThermalComfort vs ThermalPreference
IF((THERMAL_COMFORT=1)*(THERMAL_PREFERENCE="no change"),0,IF((THERMAL_COMFORT=6)*OR((THERMAL_PREFERENCE="cooler"),(THERMAL_PREFERENCE="warmer")),0,IF(OR((THERMAL_PREFERENCE="NA"),(THERMAL_COMFORT="NA")),"NA",1)))

   //%ThermalComfort vs ThermalSensation
IF((THERMAL_COMFORT=1)*(ABS(THERMAL_SENSATION)<=2),0,IF((THERMAL_COMFORT=6)*(ABS(THERMAL_SENSATION)>2),0,IF(OR((THERMAL_COMFORT="NA"),(THERMAL_SENSATION="NA")),"NA",1)))

**end for**
**Output:** Marked ASHRAE database

---

The database that has been filtered is mapped and compared with the original ASHRAE database. Fig. 5 shows the mapping with the temperature as the x-axis and relative humidity as the y-axis. The original database map is shown on the left side, where the filtered database is shown on the right side. This figure shows the data mapping based on the three TSV class groups, which are "no change", "warmer", and "cooler". The ASHRAE database is shown to have major overlapped between classes. This significant overlap is the cause of the difficulties in training the AI using this database. It is challenging to have a proper classification process with the risk of overfitting.

The class overlap is reduced with the filtered database shown on the right side. This method gives the learning process a better chance to generate better training results for proper classification. The mapping position between the "warmer" class and "cooler" class looks physiologically better than the original database. The other parameters map can be seen in Appendix C. The figure in Appendix C shows the database map for clothing insulation against the indoor temperature, the occupants' activity/metabolism against the indoor temperature and the occupants' age against the indoor temperature. Like the relative humidity and temperature map, the classes in these parameters have a better condition to be classified after the database filtering process.

This map also shows a gap in the data availability for the "warmer" class and "cooler" class. Not only to make the feature space to be equal. Furthermore, the system needs a different range of data to be registered in the database for the "warmer" and "cooler" classes. It needs more data outside the comfort temperature zone for a better learning process. The answer to this problem is semantic augmentation. The result of this method will be covered in subchapter 4.3.

### 4.1 Excluding Air Movement to Accommodate the IoT Systems and Excluding the Age

One of the six crucial parameters in thermal comfort is air movement. However, this parameter can not be easily obtained from the IoT sensor system. This work deploys the system with five parameters and four parameters without the age data of the occupant. This case study is for simulating in case that the age information of the occupants is not available. The result of both systems is compared to show their characteristics. This combination of original and filtered ASHRAE database was then used for the training data for 29 well known AI algorithms for classification. The accuracy result for each database and method can be seen in Appendix D. The parameters for each classification method used in Appendix D can be seen in Appendix E. The average of the accuracy results is given in Table 1.

The result shows that the data filtering increases the accuracy of the training results. The proposed filtering methods can make the ASHRAE database better for each RP-884 database, Thermal Comfort Database II and the combination between both databases. The accuracy increase for all methods of training. When tested against the whole original database, the result is still better than the training with the non-filtered data. Besides this data filtering, data normalisation is already included in the AI process to gain better results. The result also shows that reducing the parameter (age parameter) can maintain accuracy. This result can be caused by the overfitting or the unbalance of the feature space. The semantic data augmentation will give better accuracy results for this problem. This semantic augmentation performance will be discussed along with the psychrometric chart in subchapter 4.2.

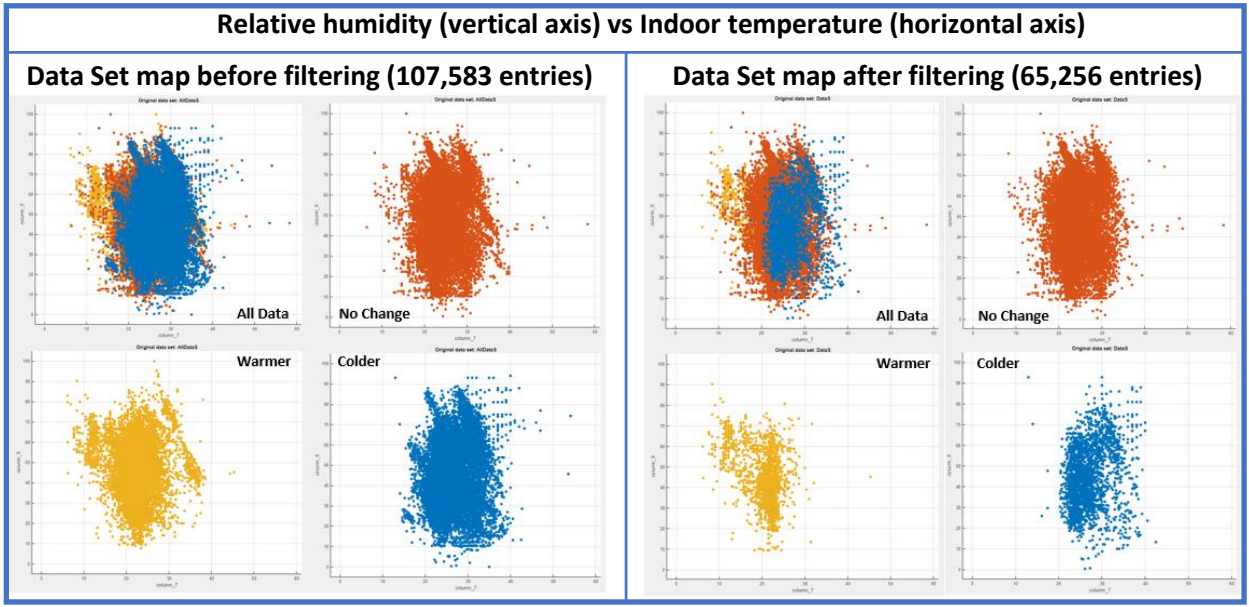

Fig. 5. The ASHRAE Database Mapping for Relative Humidity vs Indoor Temperature Before Filtering and After Filtering

TABLE 1
THE AVERAGE ACCURACY FROM 29 CLASSIFICATION METHODS

| Database | Average (%) |
|---|---|
| ASHRAE RP-884 | 51.87 |
| Filtered ASHRAE RP-884 | 76.09 |
| Filtered ASHRAE RP-884 Tested with All Data | 49.05 |
| ASHRAE RP-884 without age | 55.05 |
| Filtered ASHRAE RP-884 without age | 80.40 |
| Filtered ASHRAE RP-884 without age Tested with All Data | 50.70 |
| ASHRAE Global Thermal Comfort Db II | 42.00 |
| Filtered ASHRAE Global Thermal Comfort Db II | 70.23 |
| Filtered ASHRAE Global T C DB II Tested with All Data | 47.23 |
| ASHRAE Global T C Db II without age | 50.51 |
| Filtered ASHRAE Global T C Db II without age | 81.57 |
| Filtered ASHRAE GTC DB II without age Tested with All Data | 49.44 |
| ASHRAE RRP-884 and DB II | 43.69 |
| Filtered ASHRAE RRP-884 and DB II | 74.90 |
| Filtered ASHRAE RRP-884 and GTC DB II Tested with All Data | 48.75 |
| ASHRAE RRP-884 and G T C DB II without age | 50.53 |
| Filtered ASHRAE RRP-884 and G T C DB II without age | 81.52 |
| Filtered ASHRAE RRP-884 and GTC DB II w/o age Tested with All Data | 50.87 |

## 4.2 Validation Using Psychrometric Chart, and The Result with Data Filtering and Semantic Augmentation

Many of the previous AI works only limit the training validation with the available data. Usually, data is separated randomly into training data, testing data and validation data and the accuracy is purely based on this data. The problem raised is the edge of the comfort zone or the zone outside the predefined zone, which might still be comfortable to the occupants. This area should be explored to define the system with the ability to conserve energy. The intelligent system can have the recommender function to lower the energy cost by informing the occupants about clothing or activities that can keep them comfortable but with less energy spending. The occupants still have the probability of staying comfortable with higher clothing insulation during winter to conserve heating energy. On the contrary, the occupants will also have the probability to be comfortable in the hot summer by using lighter clothes, using the fan and consuming fresh beverages to conserve the cooling energy. This behaviour is the current gap in the previous work.

This work accommodates these needs by proposing the validation process using the psychrometric chart and test data generator. The test data generator works similar the **algorithm 1** but with the parameter range to be more specific on the comfort zone map. The temperature can be between 10°C to 40°C, with the relative humidity value between 0 to 100%. The generated data then being fed to the intelligent system, and the result is drawn in the psychrometric chart. Each label can be drawn in the chart with a different colour to show the "no change" class, "warmer" class and "cooler" class. The previous work also mapped the training result with the psychrometric chart but without the test data generation. This method makes the validation only limited to the available data, and the comfort zone can not be adequately mapped. The comfort zone can be appropriately mapped using the generated data, with the edges of the thermal comfort zone visually presented.

In order to compare the learning result and the effect of filtering and semantic augmentation method, the neural network algorithm is used in this work. As shown in Fig. 6, the neural network structure is used for the training with the original ASHRAE data, the filtered data, and the filtered semantic augmented data. The wide neural network used has 100 hidden layers. The learning dataset composition for this Neural Network training was 70% training. 15% validation and 15% testing data. The data selection for these groups is based on random selection. The parameters involved in this trial consisted of indoor temperature, relative humidity, clothing insulation, activity/metabolism and age. With an accuracy of 45.6%, the AI system is then fed with the generated test data. For this test, the parameters involved was the combination of indoor temperature, relative humidity, clothing insulation, activity/metabolism. The age parameter is a single value, taken randomly to reduce the training data size. The different age parameter values will be covered in the following subchapter, 4.3. The class result is drawn in the psychrometric chart, as shown in Fig 7 (a). This result shows an incomplete comfort zone. Only class "no change" and "cooler" classes dominate, and the comfort zone is not drawn correctly. The class "no change" is represented with green colour, "cooler" with red colour and "warmer" with blue colour. The comfort zone range is from 10°C up to 30 °C, which is not valid compared with the PMV-PPD results.

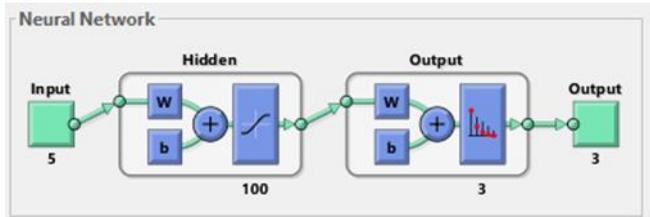

Fig. 6. Simple Neural Network Structure for Psychrometric Chart Validation.

Figure 7 (b) shows the psychrometric map of the generated test data for the system, which was trained using the filtered ASHRAE database. The parameters and dataset grouping method are similar to the trial for Fig. 7 (a). The accuracy of this training was 55.5%. All three classes are visible, but the comfort zone is still not drawn correctly. The class "warmer" only covers a small portion of the chart, and the comfort zone range spans 10°C up to more than 40 °C, which is not valid if compared with the PMV-PPD results. This problem shows that the system needs semantic augmentation to generate the correct result.

The training data with semantic augmentation was deployed with the accuracy of this training data about 98%. The parameters and the data grouping is similar to the trial for Fig. 7 (a) and Fig. 7 (b). The psychrometric map generated for the semantic augmentation filtered ASHRAE data is shown in Fig 7 (c). The comfort zone is better represented in this result. The comfort zone ranges from 17.5°C to about 29 °C. The result is better than the two mapping before, shown in Fig. 7 (a) and Fig. 7 (b). This result represents the comfort zone presented in Fig.1.

Learning from the accuracy of data training, which can be high, as shown in Appendix D, the system training result still needs to be further validated. The validation method can be the psychrometrics mapping of the comfort zone. This Fig. 7 highlights the importance of validation using a psychrometric chart

### 4.3 Parametric Visualisation in the Psychrometric Chart

The parameters shown in the previous subchapter can show the differences between classes "warmer", "no change", and "cooler" for the particular value of a parameter such as age to show the potential comfort zone for each parameter value. If the parameter is changed accordingly, the impact of this parameter change on the thermal comfort zones can be mapped and studied. The impact of the parameter change in the thermal comfort zone can be seen in Fig. 8.

This paper assesses the age parameter impact on the thermal comfort zone. This assessment becomes the example

(a)
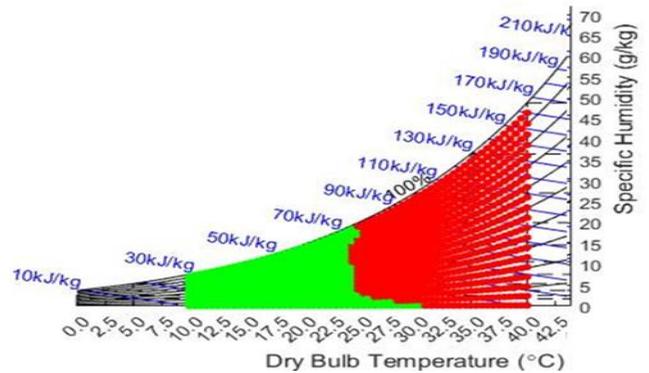

(b)
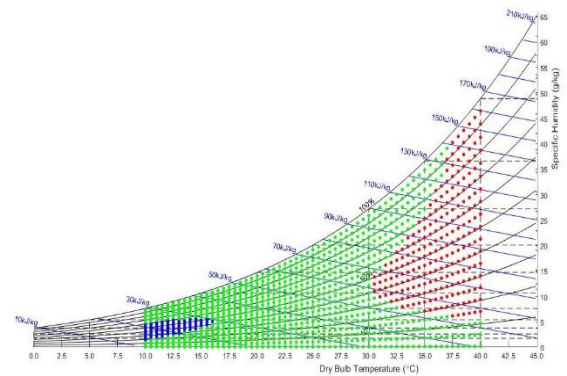

(c)
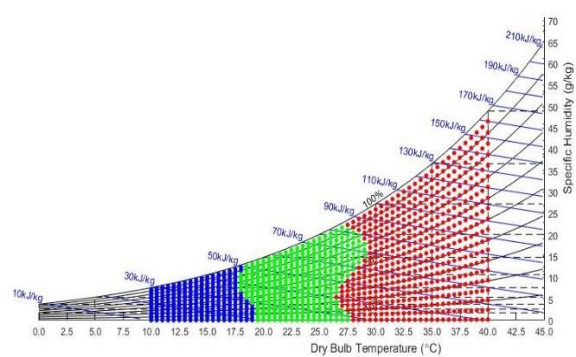

Fig. 7. Psychrometric Mapping for the Comfort Zone Trained with (a) the Original ASHRAE Database (b) the Filtered ASHRAE Database (c) the Filtered ASHRAE Database with the Data Semantic Augmentation

of the parameter assessment with this method. The age parameter is one parameter that can show the human condition aspect of thermal comfort. The thermal comfort is not standard and is based on personal factors. It has been studied that the young, elder, disabled or temporary ill people group will have a different comfort zone.

Figure 8 (a) display the representation of the comfort zone from the adult based on the filtered ASHRAE database with the data semantic augmentation. Figure 8 (b) shows the same comfort zone for the elder group. They were based on the filtered ASHRAE database with the data semantic augmentation. Based on this chart, the system designer will have insight into designing and testing the AI system since the borders between the comfort zone are not crisp and more personal.

A similar result also happened in the same age group. The differences in clothing insulation and activity/metabolism can have different thermal conditions. This condition is shown in Fig. 8 (b), 8 (c) and 8 (d). Figure 8 (b) represent class "no change", Fig. 8 (c) represent class "warmer" and Fig. 8 (d) represent class "cooler". This map shows that the class "no change" intersect with the class "warmer" and "cooler". This case means that if one person feels cold and needs a warmer environment, another person can feel comfortable. When one person feels hot and needs a cooler indoor environment, the other person can feel comfortable. The more extreme condition is the overlapping between Fig. 8 (c) and Fig 8 (d). It means that one person wanted the temperature to be warmer at the same temperature, but the other person wanted a cooler temperature. This condition highlights the need for the system to have the manual override so that the occupants can alter the system setting.

### 4.4 Potential Refinement of the Model and the Future Works

Semantic augmentation has proven to be robust in the processing of thermal data. There is the possibility that the semantic augmentation can be implemented in other parameters without changing the notion of comfort that is stored in the ASHRAE database. The humidity parameter can be one of the candidates for future work. There is no comfort recommendation for the humidity value for comfort, but the healthy range for the relative humidity is not more than 80% and not less than 15%. This gap can lead to registering the semantic augmentation for the relative humidity comfort value. This system can be developed to control the indoor humidifier or the dehumidifier.

Another potential development of the system is implementing the recommendation and gamification system to lower energy use but maintain comfort. Since thermal comfort is the state of mind related to memory and not just physiology, the gamification feature and the intelligent system preset can help achieve the goal of lower energy use either for heating or cooling. The system can influence the user to feel comfortable with the gamification and recommendation, but it will need a long adaptation process [8]. For the low temperature, for example, exposure to cold acclimation can improve the subjective responses to cold [25]. This is why this research for the use of the ASHRAE database is essential, to give the fundamental ability to the intelligent system to deliver comfort. There can also be a healthier target set in the system, like exposing the user to lower temperatures to decrease body fat [23].

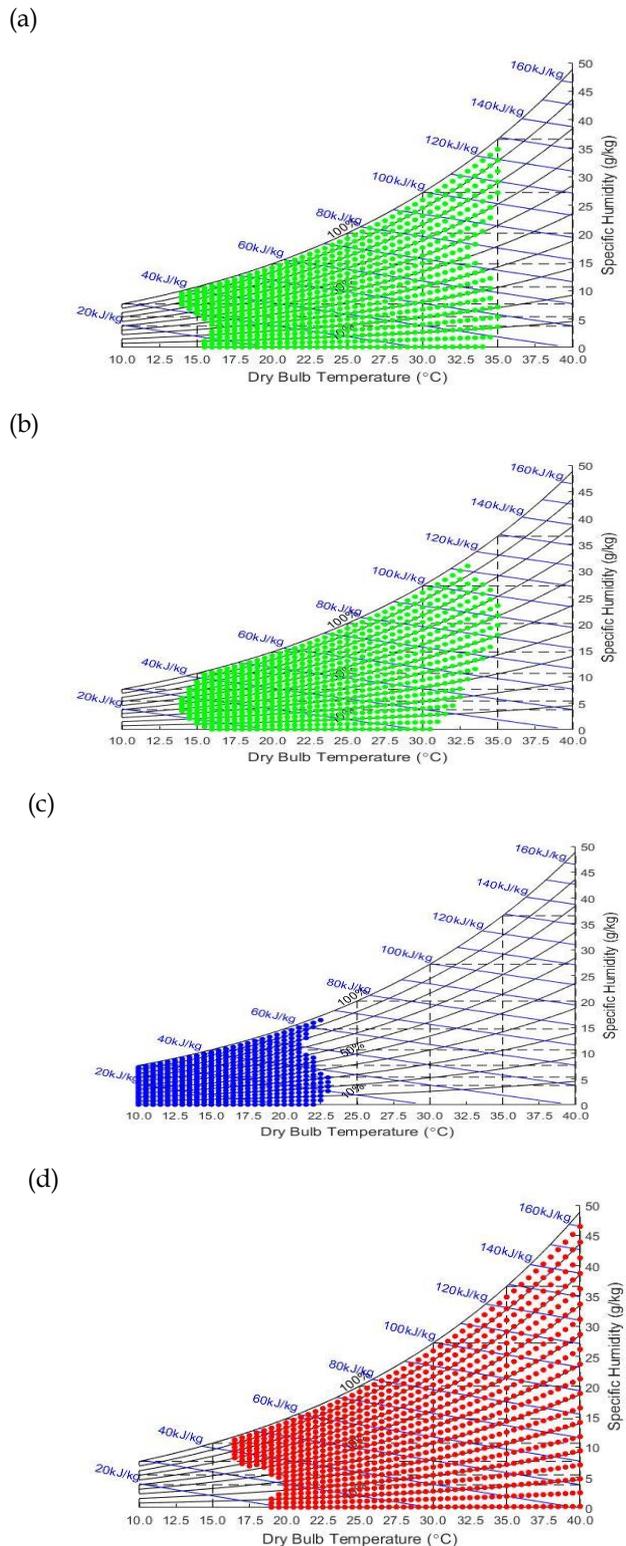

Fig. 8. Psychrometric Mapping for (a) the Comfort Zone Trained with the Filtered ASHRAE Database with the Data Semantic Augmentation for adult, (b) for elderly people group (c) the "warmer" class of the elder people group (d) the "cooler" class of the elderly people group

# 5 CONCLUSION

Obtaining the dataset for thermal comfort is not an easy task. This work develops the filtering and semantic augmentation for the ASHRAE database, one of the most reliable databases for thermal comfort. This work also proves that the database can perform well in the thermal comfort zone prediction. This work shows that even the training for the AI process has been done with an excellent training validation percentage, it does not guarantee that the system will perform well for the data with extreme value or within the comfort zone borders. In line with that findings, this work also proposes validation methods based on the test data generation and validation through psychrometric comfort zone mapping. This method will help analyse each impact of the parameters for thermal comfort based on the psychrometric mapping of the thermal comfort zone.

**Kanisius Karyono** currently is pursuing his PhD in the department of the Built Environment, Liverpool John Moores University. He received his bachelor in Informatics from Atma Jaya Yogyakarta University in 2000 and a bachelor in Electrical Engineering from Gadjah Mada University Indonesia in 2002. He received his Master in Electrical Engineering from Pelita Harapan University Indonesia in 2009. He works at the Department of Electrical Engineering, Universitas Multimedia Nusantara Indonesia. Kanisius researches Human Comfort (Lighting and Thermal Comfort), Internet of Things, Smart System, Wireless Sensor Network, communication protocol modelling, and Industrial 4.0.

**Badr Abdullah** is the Programme Leader of Architectural Engineering & Building Services Engineering at the Department of the Built Environment. Badr gained an MEng in Electrical Engineering and Electronics in 2004 and a PhD in Vision Monitoring Systems for Hybrid Welding Processes and Hazardous Environments in Real-time in 2007 from the University of Liverpool. Badr's main teaching areas are Engineering and Applied Mathematics, Engineering Principles and Electrical Engineering. As well as teaching, Badr heads the i4Sens3D (Industry 4, Sensors and Drones) Research Group in the Faculty of Engineering and Technology. Badr has extensive research interests spanning a wide range of unique applied science and technologies for the design and development of sensors to suit a wide range of industries such as built environment, energy, oil/gas, environmental monitoring, manufacturing and healthcare.

**Alison Cotgrave** is the Professor of Built Environment Education and Subject Leader in the School of Civil Engineering and Built Environment with responsibility for Civil Engineering, Architectural Engineering, Building Services Engineering, Construction Management, Architectural Technology, Real Estate, Quantity Surveying and Building Surveying. However, She is currently seconded to the post of Academic Registrar until May 2022. She has worked in academia for 27 years and was previously employed in the construction industry as a site manager. After graduating with BSc(Hons) Building in 1988 she subsequently completed a Masters of Education at the University of Manchester in 1997, followed by a PhD at LJMU in 2008. She is a Fellow of the Chartered Institute of Building, Member of the Royal Institution of Chartered Surveyors, Chartered Environmentalist, Fellow of the Leadership Foundation and a Senior Fellow of the Higher Education Academy. She has co-authored a number of books on Construction Technology and have published extensively in the area of Sustainability in Construction Education. She is currently on the supervision teams for a number of PhD students whose topics range from construction pedagogy to the design of a decision making aid for the selection of renewable/sustainable energy systems for buildings. She is the External Examiner at a number of HEIs and a member of the Chartered Institute of Building accreditation panel, which reviews and accredits Construction Management programmes in the UK and internationally.

**Ana Bras** is a Civil Engineer, CEng, FICE, with 17 years of experience in R&D and Consultancy on bio-based composites for durability and hygrothermal risks minimisation of retrofitting solutions for buildings and RC structures. She is a Reader (Associate Professor) in Bio-materials for infrastructure at the Department of Civil Engineering and Built Environment, Liverpool John Moores University (UK). She graduated in 2004 with a MEng of a five years degree university programme in Civil Engineering at NOVA University of Lisbon. In 2011, she completed her PhD studies in Civil Engineering at NOVA (speciality in Building Rehabilitation). She is a researcher and research project manager, with active participation in several R&D projects and consultancy for institutions including: NOVA University of Lisbon (Portugal), University of Bath (UK), University of Campinas (UNICAMP, Brazil), Polytechnic Institute of Setúbal (IPS, Portugal), Institute of Welding and Quality (ISQ, Portugal) and more recently, at LJMU. Concurrently, she is taking part in HORIZON 2020 task groups to evaluate EU projects. Ana develops research work in the field of performance based design with bio-inspiring solutions (laboratory and computa-tional-based) for the selection of the best retrofitting solutions for buildings and structures. She uses probabilistic approaches to increase the effectiveness of sustainable solutions in terms of their service life, and to estimate energy savings and hygrothermal risks.

**Jeff Cullen** graduated from the University of Liverpool, School of Electrical Engineering and Electronics, with a BEng(Hons) in Computer Electronics and Robotics. He then obtained a PhD at Liverpool in 2000 entitled 'Optical Techniques for Industrial Automated Applications'. His post doctoral research career began at Liverpool working for Prof. Jim Lucas on a tyre pressure monitoring system for trucks. He also worked on a novel UV microwave system, before working on an online spot welding monitor, predicting weld quality and monitoring energy usage. In 2005 he joined LJMU as a founding member of the RF and Microwave group under Prof. Ahmed Al-Shamma'a, which later became part of the newly formed BEST (Built Environment and Sustainable Technologies) Research Institute. Here he continued research related to energy, investigating the performance of fridges for the leisure industry. He has also used microwave systems to provide a novel way of heating showers and remediation of waste oil for generator and biodiesel use. He has also lectured at both undergraduate and postgraduate level, delivering courses on Automotive Electronics, Programming and Software Engineering for the School of Engineering. He has helped deliver modules related to Energy Management and Building Monitoring Systems for the School of the Built Environment.

Appendix A: Missing data for the five dominant parameters in the ASHRAE database

| Dataset Name | Number of Entry | Missing Data Entry | | | | | | | | | |
|---|---|---|---|---|---|---|---|---|---|---|---|
| | | Temperature | Temperature in % | Humidity | Humidity in % | Clothing | Clothing in % | Age | Age in % | Metabolic Rate | Metabolic Rate in % |
| ASHRAE RP-884 | 25,616 | 3816 | 14.90 | 761 | 2.97 | 332 | 1.30 | 6899 | 26.93 | 2164 | 8.45 |
| ASHRAE Global Thermal Comfort Database II | 81,967 | 3856 | 4.70 | 9060 | 11.05 | 7588 | 9.26 | 57105 | 69.67 | 15000 | 18.30 |
| TOTAL | 107,583 | 7672 | 7.13 | 9821 | 9.13 | 7920 | 7.36 | 64004 | 59.49 | 17164 | 15.95 |

Appendix B: The number of data entries filtered in each filtering item in the ASHRAE database

| Dataset Name | Number of Entry | Inconsistency 1 | % Inconsistency 1 | Inconsistency 2 | % Inconsistency 2 | Inconsistency 3 | % Inconsistency 3 | Inconsistency 4 | % Inconsistency 4 | Inconsistency 5 | % Inconsistency 5 |
|---|---|---|---|---|---|---|---|---|---|---|---|
| ASHRAE RP-884 | 25,616 | 3051 | 11.91 | 1564 | 6.11 | 10192 | 39.79 | 192 | 0.75 | 129 | 0.50 |
| ASHRAE Global Thermal Comfort Database II | 81,967 | 14236 | 17.37 | 10393 | 12.68 | 27341 | 33.36 | 1106 | 1.35 | 326 | 0.40 |
| TOTAL | 107,583 | 17287 | 16.07 | 11957 | 11.11 | 37533 | 34.89 | 1298 | 1.21 | 455 | 0.42 |

Filtering items:
(1)Thermal Acceptability vs Thermal Preference
(2)Thermal Sensation vs Thermal Acceptability
(3)Thermal Sensation vs Thermal Preference
(4)Thermal Comfort vs Thermal Preference
(5)Thermal Comfort vs Thermal Sensation

After filtering, the amount of data in ASHRAE RRP-884: 14,970 (filtered value: 10,646 or 41.56%)
After filtering, the amount of data in ASHRAE Thermal Comfort Database II: 50,286 (filtered value: 31,681 or 38.65%)
The Total data in both data set are 65,256 (filtered value: 42,327 or 39.34%)

Appendix C: Data Mapping Before and After Filtering Process

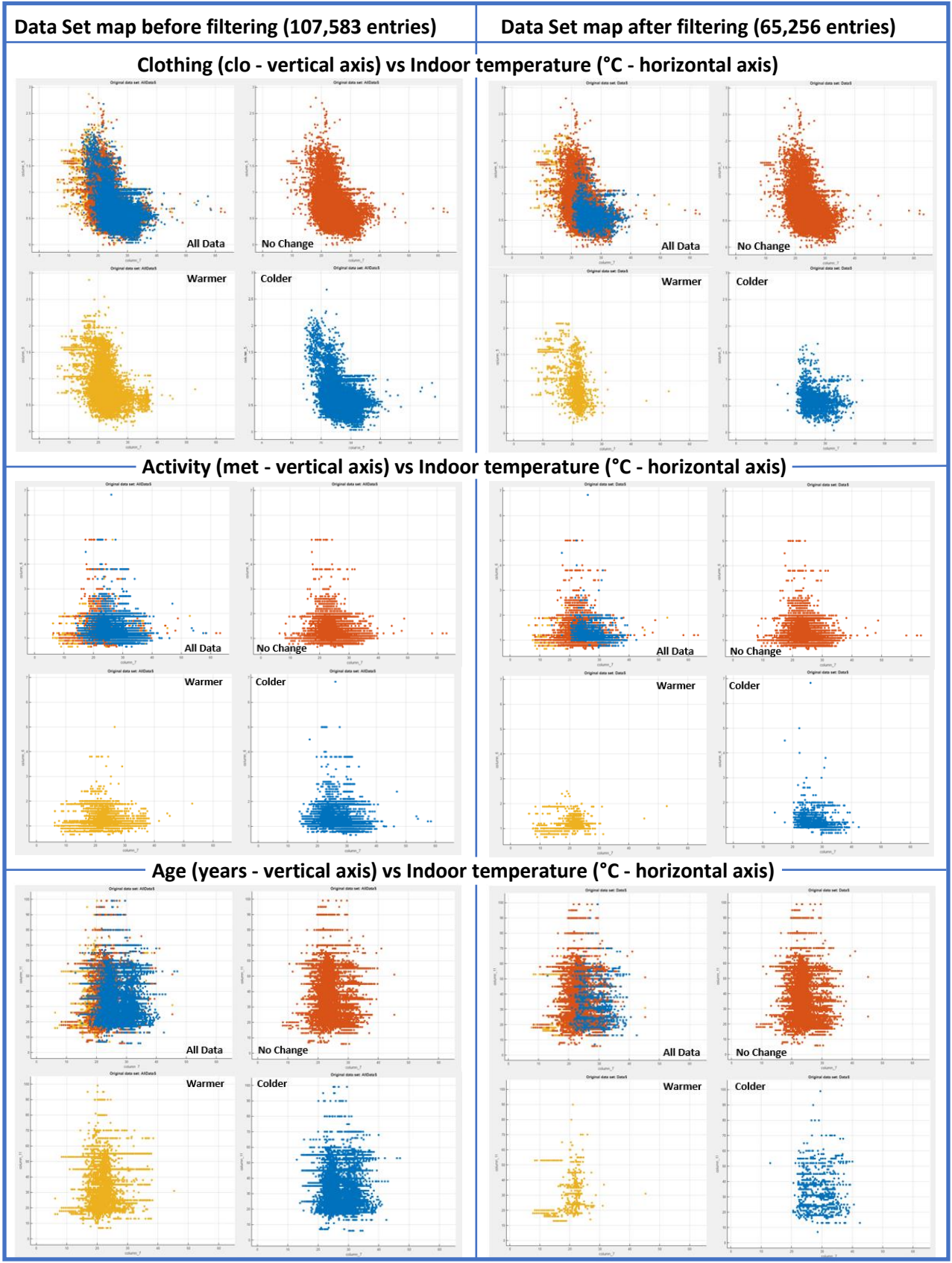

Appendix D: CLASSIFICATION METHODS ACCURACY COMPARISON FOR DATABASE ALLOCATED FOR TRAINING AND VALIDATION (IN %)

| Classification Methods | ASHRAE RP-884 | Filtered ASHRAE RP-884 | Filtered ASHRAE RP-884 Tested with All Data | ASHRAE RP-884 without age | Filtered ASHRAE RP-884 without age | Filtered ASHRAE RP-884 without age Tested with All Data | ASHRAE Global Thermal Comfort Db II | Filtered ASHRAE Global Thermal Comfort Db II | Filtered ASHRAE Global T C Db II Tested with All Data | ASHRAE Global T C Db II without age | Filtered ASHRAE Global T C Db II without age | Filtered ASHRAE GTC DB II without age Tested with All Data | ASHRAE RRP-884 and DB II | Filtered ASHRAE RRP-884 and DB II | Filtered ASHRAE RRP-884 and GTC DB II Tested with All Data | ASHRAE RRP-884 and G T C DB II without age | Filtered ASHRAE RRP-884 and G T C DB II without age | Filtered ASHRAE RRP-884 and GTC DB II w/o age Tested with All Data |
|---|---|---|---|---|---|---|---|---|---|---|---|---|---|---|---|---|---|---|
| Fine Tree | 59.20 | 89.60 | 53.90 | 60.70 | 89.40 | 53.80 | 42.50 | 90.20 | 51.60 | 48.50 | 90.30 | 51.90 | 45.00 | 89.80 | 52.40 | 49.00 | 89.7 | 52.4 |
| Medium Tree | 57.70 | 88.00 | 52.10 | 57.80 | 87.90 | 52.10 | 41.10 | 89.90 | 51.30 | 45.40 | 90.00 | 51.70 | 43.60 | 89.30 | 52.00 | 46.50 | 89.3 | 52 |
| Coarse Tree | 53.70 | 87.20 | 51.40 | 54.70 | 87.50 | 52.30 | 39.70 | 89.80 | 50.80 | 41.70 | 89.80 | 50.80 | 42.30 | 89.10 | 51.90 | 42.30 | 88.9 | 51.4 |
| Linear Discriminant | 51.40 | 87.30 | 52.00 | 52.00 | 86.80 | 52.20 | 38.50 | 89.80 | 51.30 | 40.70 | 89.50 | 52.30 | 42.00 | 89.00 | 51.60 | 43.60 | 88.7 | 52.3 |
| Quadratic Discriminant | 49.50 | 86.70 | 52.10 | 52.80 | 86.20 | 52.50 | 36.80 | 89.30 | 51.50 | 42.50 | 88.20 | 52.50 | 37.80 | 88.50 | 52.00 | 43.70 | 87.5 | 52.7 |
| Gaussian Naïve Bayes | 52.90 | 85.10 | 53.30 | 52.30 | 85.20 | 53.20 | 43.30 | 88.00 | 53.50 | 42.70 | 88.00 | 53.50 | 43.60 | 86.80 | 53.60 | 43.30 | 86.7 | 53.6 |
| Kernel Naïve Bayes | 55.70 | 87.50 | 52.30 | 55.70 | 87.30 | 52.20 | 43.20 | 90.20 | 51.90 | 42.60 | 90.10 | 51.80 | 47.40 | 89.30 | 52.30 | 46.00 | 89.1 | 52.2 |
| Linear SVM | 36.70 | 50.10 | 41.50 | 41.00 | 67.50 | 46.80 | 36.20 | 38.10 | 40.90 | 43.90 | 73.80 | 46.20 | 35.40 | 41.50 | 41.40 | 42.90 | 71.4 | 46.2 |
| Quadratic SVM | 38.20 | 50.30 | 41.50 | 35.10 | 67.50 | 46.80 | 36.40 | 38.10 | 41.10 | 37.80 | 72.50 | 46.60 | 33.40 | 41.60 | 41.40 | 29.90 | 59.1 | 45.9 |
| Cubic SVM | 27.10 | 47.20 | 41.40 | 21.00 | 48.90 | 44.50 | 26.70 | 37.80 | 41.10 | 21.70 | 32.10 | 32.20 | 24.80 | 36.50 | 41.20 | 28.00 | 62.8 | 46.7 |
| Fine Gaussian SVM | 45.40 | 53.60 | 42.50 | 52.10 | 70.50 | 48.20 | 41.20 | 38.70 | 41.50 | 56.10 | 74.80 | 46.90 | 41.40 | 42.70 | 42.40 | 53.60 | 73.2 | 47.8 |
| Medium Gaussian SVM | 38.50 | 51.80 | 42.40 | 45.20 | 68.70 | 47.60 | 38.70 | 38.20 | 41.20 | 49.70 | 74.00 | 46.40 | 38.10 | 41.90 | 41.80 | 47.40 | 72.2 | 47.1 |
| Coarse Gaussian SVM | 37.40 | 50.30 | 41.50 | 42.50 | 67.50 | 46.80 | 36.90 | 38.10 | 40.90 | 46.00 | 73.80 | 46.20 | 36.50 | 41.60 | 41.50 | 44.00 | 71.8 | 46.6 |
| Fine KNN | 73.00 | 95.60 | 53.10 | 87.20 | 97.90 | 53.60 | 50.70 | 91.10 | 51.90 | 79.30 | 97.10 | 55.90 | 56.00 | 92.40 | 54.40 | 81.20 | 97.4 | 58.8 |
| Medium KNN | 57.40 | 89.30 | 52.50 | 62.50 | 89.10 | 52.90 | 43.70 | 90.00 | 51.20 | 55.60 | 90.50 | 52.10 | 46.70 | 89.60 | 52.30 | 56.20 | 89.9 | 53 |
| Coarse KNN | 54.20 | 87.80 | 52.30 | 57.70 | 87.70 | 52.10 | 42.20 | 89.80 | 50.80 | 49.70 | 90.00 | 51.00 | 44.50 | 89.00 | 51.30 | 50.50 | 89.2 | 51.6 |
| Cosine KNN | 56.60 | 88.90 | 52.50 | 60.90 | 88.40 | 52.60 | 43.40 | 89.90 | 51.10 | 53.80 | 90.30 | 51.80 | 46.30 | 89.40 | 52.10 | 54.50 | 89.4 | 52.5 |
| Cubic KNN | 57.30 | 89.30 | 52.50 | 62.20 | 89.00 | 53.00 | 43.70 | 90.00 | 51.20 | 55.50 | 90.50 | 52.20 | 46.70 | 89.60 | 52.30 | 56.10 | 89.9 | 53 |
| Weighted KNN | 73.00 | 95.60 | 53.60 | 87.30 | 97.90 | 54.30 | 50.90 | 91.10 | 51.60 | 70.80 | 97.20 | 55.00 | 56.10 | 92.40 | 53.80 | 81.60 | 97.4 | 57.3 |
| Ensemble Boosted Trees | 58.30 | 88.30 | 52.10 | 58.20 | 88.10 | 52.20 | 41.60 | 89.90 | 51.10 | 45.70 | 90.00 | 51.00 | 43.70 | 89.30 | 51.60 | 47.30 | 89.3 | 51.8 |
| Ensemble Bagged Trees | 78.10 | 95.50 | 53.70 | 88.20 | 97.60 | 54.80 | 53.70 | 91.10 | 51.70 | 80.40 | 96.60 | 54.90 | 56.70 | 92.30 | 53.90 | 81.90 | 96.9 | 57 |
| Ensemble Subspace Discriminant | 52.70 | 87.20 | 52.10 | 52.70 | 87.00 | 51.90 | 43.90 | 90.00 | 51.10 | 41.70 | 90.10 | 51.20 | 44.90 | 89.20 | 51.90 | 43.80 | 89 | 51.6 |
| Ensemble Subspace KNN | 94.10 | 99.10 | 53.50 | 78.50 | 94.80 | 53.30 | 79.80 | 97.60 | 56.40 | 59.50 | 93.80 | 53.30 | 86.30 | 98.60 | 58.60 | 58.50 | 93.2 | 54.4 |
| Ensemble RUS Boosted Trees | 41.20 | 49.20 | 47.00 | 42.90 | 57.00 | 47.20 | 41.10 | 57.20 | 50.10 | 42.40 | 43.80 | 47.10 | 36.30 | 48.90 | 47.70 | 35.00 | 46.7 | 48.5 |
| Narrow Neural Network | 40.00 | 88.50 | 52.50 | 45.90 | 68.50 | 47.80 | 38.00 | 35.50 | 31.90 | 47.30 | 73.10 | 43.40 | 37.30 | 89.20 | 42.00 | 45.90 | 71.4 | 44 |
| Medium Neural Network | 40.80 | 52.30 | 42.50 | 46.90 | 69.00 | 48.40 | 38.40 | 37.90 | 41.60 | 49.50 | 74.10 | 46.80 | 37.80 | 89.40 | 42.10 | 47.20 | 72.2 | 47.2 |
| Wide Neural Network | 42.20 | 90.40 | 52.50 | 50.10 | 70.10 | 48.60 | 34.60 | 36.20 | 32.20 | 49.30 | 90.30 | 52.00 | 38.80 | 42.20 | 42.60 | 49.40 | 72.5 | 47.9 |
| Bilayered Neural Network | 40.50 | 88.80 | 52.50 | 48.50 | 53.10 | 48.20 | 38.20 | 89.70 | 51.30 | 49.50 | 73.30 | 44.10 | 37.80 | 89.30 | 51.90 | 47.80 | 89.2 | 51.8 |
| Trilayered Neural Network | 55.50 | 52.30 | 42.50 | 48.60 | 89.10 | 52.80 | 26.80 | 89.70 | 51.30 | 50.50 | 90.20 | 51.70 | 38.00 | 89.40 | 52.30 | 48.20 | 89.5 | 52.3 |
| **Average** | **52.36** | **77.68** | **49.49** | **55.21** | **81.40** | **50.95** | **41.79** | **72.86** | **47.76** | **49.65** | **82.68** | **49.74** | **43.62** | **76.82** | **49.18** | **49.84** | **82.53** | **51.02** |

The Detailed Parameters of Each Method is presented in Appendix E

Appendix E: Parameters for Each Classification Methods

| Fine Tree | Gini's diversity index, 100 max number of splits |
|---|---|
| Medium Tree | Gini's diversity index, 20 max number of splits |
| Coarse Tree | Gini's diversity index, 4 max number of splits |
| Linear Discriminant | Full covariance structure |
| Quadratic Discriminant | Full covariance structure |
| Gaussian Naïve Bayes | Gaussian, unbounded support |
| Kernel Naïve Bayes | Gaussian Kernel type, unbounded support |
| Linear SVM | 1 box constraint level, auto kernel scale mode, one-vs-one multiclass method, standardise data |
| Quadratic SVM | 1 box constraint level, auto kernel scale mode, one-vs-one multiclass method, standardise data |
| Cubic SVM | 1 box constraint level, auto kernel scale mode, one-vs-one multiclass method, standardise data |
| Fine Gaussian SVM | 1 box constraint level, kernel scale: 0.56, one-vs-one multiclass method, standardise data |
| Medium Gaussian SVM | 1 box constraint level, kernel scale: 2.2, one-vs-one multiclass method, standardise data |
| Coarse Gaussian SVM | 1 box constraint level, kernel scale: 8.9, one-vs-one multiclass method, standardise data |

| Model | Parameters |
|---|---|
| Fine KNN | 1 Number of neighbours, Euclidean distance metric, Equal distance weight, standardise data |
| Medium KNN | 10 Number of neighbours, Euclidean distance metric, Equal distance weight, standardise data |
| Coarse KNN | 100 Number of neighbours, Euclidean distance metric, Equal distance weight, standardise data |
| Cosine KNN | 10 Number of neighbours, Cosine distance metric, Equal distance weight, standardise data |
| Cubic KNN | 10 Number of neighbours, Minkowski (Cubic) distance metric, Equal distance weight, standardise data |
| Weighted KNN | 10 Number of neighbours, Euclidean distance metric, Squared distance weight, standardise data |
| Ensemble Boosted Trees | AdaBoost ensemble method, Decision Tree learner type, 20 maximum number of splits, 30 number of learners, 0.1 learning rate |
| Ensemble Bagged Trees | Bag ensemble method, Decision Tree learner type, 65260 maximum number of splits, 30 number of learners |
| Ensemble Subspace Discriminant | Subspace ensemble method, Discriminant learner type, 30 number of learners, 3 Subspace dimension |
| Ensemble Subspace KNN | Subspace ensemble method, Nearest neighbours learner type, 30 number of learners, 3 Subspace dimension |
| Ensemble RUS Boosted Trees | RUSBoost ensemble method, Decision Tree learner type, 20 maximum number of splits, 30 number of learners, 0.1 learning rate |
| Narrow Neural Network | 1 Number of fully connected layers, layer size 10, ReLU Activation, 1000 iteration limit, 0 regularisation strength (Lambda), standardise data |
| Medium Neural Network | 1 Number of fully connected layers, layer size 25, ReLU Activation, 1000 iteration limit, 0 regularisation strength (Lambda), standardise data |
| Wide Neural Network | 1 Number of fully connected layers, layer size 100, ReLU Activation, 1000 iteration limit, 0 regularisation strength (Lambda), standardise data |
| Bilayered Neural Network | 2 Number of fully connected layers, first and second layer size 10, ReLU Activation, 1000 iteration limit, 0 regularisation strength (Lambda), standardise data |
| Trilayered Neural Network | 3 Number of fully connected layers, first, second and third layer size 10, ReLU Activation, 1000 iteration limit, 0 regularisation strength (Lambda), standardise data |